\documentclass{elsart}

\usepackage{amsmath}

\usepackage{graphicx}

\usepackage{bm}

\journal{Physics Letters  A}
\begin{document}

\begin{frontmatter}

\title{Bias driven coherent carrier dynamics in a two-dimensional
aperiodic potential}

\author{F.\ A.\ B.\ F.\ de Moura}
\address{Instituto de  F\'{\i}sica, Universidade Federal de
Alagoas, Macei\'{o} AL 57072-970, Brazil}

\author{L.~P.~ Viana}
\address{P\'olo Penedo, Universidade Federal de Alagoas, Penedo
AL 57200-000, Brazil}

\author{M.~L.~Lyra}
\address{Instituto de  F\'{\i}sica, Universidade Federal de
Alagoas, Macei\'{o} AL 57072-970, Brazil}

\author{V.~A.~ Malyshev}
\thanks{On leave from V. A. Fock Institute of Physics, St.~Petersburg
University, 198904 St.-Petersburg, Russia}
\address{Centre for Theoretical Physis and Zernike Institute for Advanced
Materials, University of Groningen, Nijenborgh 4, 9747 AG Groningen, The
Netherlands}

\author{F.~Dom\'{\i}nguez-Adame}
\address{GISC, Departamento de F\'{\i}sica de Materiales,
Universidad Complutense, E-28040 Madrid, Spain}

\begin{abstract}
We study the dynamics of an electron wave-packet in a
two-dimensional square lattice with an aperiodic site potential in
the presence of an external uniform electric field. The aperiodicity
is described by  $\epsilon_{\bf m} = V\cos{(\pi\alpha
m_x^{\nu_x})}\cos{(\pi\alpha m_y^{\nu_y})}$ at lattice sites $(m_x,
m_y)$, with $\pi \alpha$ being a rational number, and $\nu_x$ and
$\nu_y$ tunable parameters, controlling the aperiodicity. Using an
exact diagonalization procedure and a finite-size scaling analysis,
we show that in the weakly aperiodic regime ($\nu_x,\nu_y < 1$), a
phase of extended states emerges in the center of the band at zero
field giving support to a macroscopic conductivity in the
thermodynamic limit. Turning on the field gives rise to Bloch
oscillations of the electron wave-packet. The spectral density of
these oscillations may display a double peak structure signaling
the spatial anisotropy of the potential landscape. The frequency of
the oscillations can be understood using a semi-classical approach.

\end{abstract}

\end{frontmatter}

\section{Introduction}

Materials with restricted geometry, such as semiconductor
quantum-well structures,~\cite{Khitrova99} quantum dots and
wires,~\cite{QW,Brandes05} organic thin films~\cite{Tischler07} as
well as quasiperiodic structures,~\cite{ELA} are nowadays
subjects of growing interest from both fundamental and practical
points of view. An attributive peculiarity of almost all of them is
the presence of disorder, which can be both of an intrinsic nature
(imperfections of the structure itself) and originated from a
random environment.

Whenever disorder is involved, Anderson's ideas about localization
of quasiparticles states come into play.~\cite{Anderson58} In three
dimensions (3D), the states at the center of the quasiparticle band
remain extended for a relatively weak disorder (of magnitude smaller
than the bandwidth), while the other states (in the neighborhood of
the band edges) turn out to be exponentially localized. This implies
the existence of two mobility edges which separates the phases of
extended and localized states.~\cite{Mott68} On the contrary,
uncorrelated disorder of any magnitude causes localization of all
one-particle eigenstates in one dimension (1D)~\cite{Mott61} and two
dimensions (2D).~\cite{abrahams}

Since late eighties, however, it has been realized that extended
states may survive on 1D systems if the disorder distribution is
correlated.~\cite{Flores89,Dunlap90,phillips91,adame1,adame2,%
Bellani99,Moura98,Izrailev99,Kuhl00} Thus, a short-range correlated
disorder was found to stabilize the extended states at special resonance
energies. In the thermodynamic limit, such extended states form a
set of null measure in the density of states,~\cite{Flores89,Dunlap90,%
phillips91,adame1,adame2} implying the absence of mobility edges in
these systems. In contrast, systems with long-range correlations of
disorder support a set of delocalized
states within a finite bandwidth,~\cite{Moura98,Izrailev99}
giving rise to mobility edges.
Theoretical predictions of localization suppression on 1D
geometries, due to correlations of the disorder distribution, were
confirmed experimentally in semiconductor superlattices
with intentional correlated disorder,~\cite{Bellani99} as well as in
single-mode waveguides with correlated scatterers.~\cite{Kuhl00}

Among 1D models with extended states, aperiodic Anderson
models~\cite{Sarma88Sarma90} with an incommensurate site potential
represent a class of particular interest. These models have been
extensively investigated in the
literature,~\cite{Grempel82,Griniasty88,Thouless88,Sarma88Sarma90,%
Yamada04} and the localized or extended nature of the eigenstates
has been related to general characteristics of aperiodic site-energy
distributions. Interesting realizations of this kind of distribution
are achieved when weakly aperiodic potentials are considered: these
systems support  the occurrence of mobility edges, in a close
analogy with the standard 3D Anderson model.~\cite{Sarma88Sarma90}

A key question in any general theory of aperiodic systems is the
relationship between a given aperiodic site-energy distribution and
physical properties of the system. Nowadays, the solution of this
problem is still lacking.~\cite{Macia2006} Nevertheless, some
important results have been obtained concerning the dynamics of
elementary excitations, like electrons, phonons, excitons,
polaritons, spin waves, plasmons or magnons, propagating through
different classes of aperiodic
lattices.~\cite{Macia2000,ELA,deMoura2006,chicomag2007} In these
studies, relevant physical properties were analyzed in terms of
appropriate model Hamiltonians.

Recently, the dynamics of a single electron in homogeneous
systems subjected to
a uniform static electric field has received much attention. Under
this constraint, the electron undergoes the so-called Bloch
oscillations,~\cite{Bloch28,Zener34,Esaki70,Wannier,Dunlap86} whose
amplitude is proportional to the energy bandwidth.
Electronic Bloch oscillations were observed for the first time in
semiconductor superlattices~\cite{Leo92}  (see
Ref.~\cite{Leo98} for an overview). A similar phenomenon of
sustained oscillations of the electromagnetic field, named photon
Bloch oscillations, have been also found in 2D waveguide arrays and
optical superlattices based on porous silicon.~\cite{Gil04} In
disordered systems, Bloch oscillations were predicted to occur in
both 1D~\cite{prl03} and 2D~\cite{JPCMat07} geometries whenever the
disorder distribution is long-range correlated. The amplitude of the
oscillations was found to carry information about the energy
difference between the two mobility edges, existing in these
systems. The occurrence of Bloch oscillations was also theoretically
argued for 1D chains with aperiodic slowly varying
potentials.~\cite{prbsarma} A spectroscopic manifestation of Bloch
oscillations is found in the appearance of the Wannier-Stark ladder
in the absorption spectrum.~\cite{Diaz06}

In this paper, apart from the generalization of the previous
work,~\cite{Sarma88Sarma90} done for a one-dimensional system, to a
2D system, we consider an electron subjected to an external uniform
electric field. The latter brings new features to the electron wave
packet dynamics moving in incommensurate potential. In particular,
despite a quasi-stochasticy of the system, sustained Bloch
oscillations can be observed, previously believed to exist only for
a periodic structures. In order to produce  an aperiodic (2D) site
potential, the formalism used in Ref.~\cite{Sarma88Sarma90}
was considered. It consists in using a sinusoidal function whose
phases $\phi_x$ and $\phi_y$ varies as a power-law. The power-law
exponent controls the degree of aperiodicity in the site potential.
We start by employing an exact diagonalization formalism to compute
the {\it dc} conductance and participation number in the absence of
the electric field. We numerically demonstrate that their size
behavior is consistent with the existence of extended states near
the band center for weakly aperiodic  potentials. In this regime,
the wave-packet dynamics becomes ballistic. Furthermore, we focus on
the wave-packet dynamics in the presence of an external uniform
electric field. We show that in the limit of a weakly aperiodic site
potential, the electric field promotes sustained Bloch oscillations
of the electronic wave-packet. The frequency of these oscillations
can be understood within the framework of a semi-classical approach.

The outline of the paper is as follows. In the next section we
present our model and quantities of interest. In Sec.~\ref{Results},
the results of the numerical simulations of the dc conductance,
participation number and electronic wave-packet dynamics are
discussed. We summarize in Sec.~\ref{Summary}.

\section{Model and formalism }
\label{Model}

We consider a tight-binding single electron Hamiltonian on a regular
2D open lattice of spacing $a$ with an aperiodic site potential
and a uniform static electric field~\cite{Dunlap86,Nazareno99}
\begin{eqnarray}
H & = & \sum_{\bm m} \left(\epsilon_{\bm m}+\bm{U}\cdot \bm{m}
\right) |{\bm m}\rangle\langle {\bm m}|  \nonumber \\
&+& J\sum_{\langle \bm{mn}\rangle} \Big(|{\bm m}\rangle\langle{\bm n}|
+ |{\bm n}\rangle\langle{\bm m}|\Big) \ ,
\label{hamiltonian}
\end{eqnarray}
where  $|{\bm m}\rangle$ is a Wannier state localized at site
$\bm{m}=m_x{\bm e}_x+m_y{\bm e}_y$, $\epsilon_{\bm m}$ is its
energy, and $\bm{U} = e\bm{F}a$ is the energetic bias given
by the electric field $\bm{F}$, $-e$ being the electron charge. Here
${\bm e}_x$ and ${\bm e}_y$ are the corresponding Cartesian unit
vectors. We will assume that the electric field $\bm{F}$ is applied
along the diagonal of the square lattice, i.e., $\bm{U} =
U({\bm e}_x+{\bm e}_y)/\sqrt{2}$, were $U$ is the bias.
Transfer integrals are restricted to the nearest-neighbor
interaction and are given by $J$. Hereafter, the energy scale is
fixed by setting $J=1$.

The 2D aperiodic potential $\epsilon_{\bm m}$ is taken in the form:
\begin{eqnarray}
\epsilon_{\bm m}=V \cos{(\pi \alpha m_x^{\nu_x})}\cos{(\pi\alpha
m_y^{\nu_y})}\ , \label{aperiodic}
\end{eqnarray}
where $V$, $\alpha$, $\nu_x$, and $\nu_y$ are variable parameters.

Equation~(\ref{aperiodic}) is a 2D generalization of the aperiodic
1D potential introduced by Das Sarma {\it et
al}.~\cite{Sarma88Sarma90} In $1D$ and at $\nu=1$, this potential
represents just Harper's model. In this case, a rational $\alpha$
describes a crystalline solid, whereas an irrational $\alpha$ yields
an incommensurate potential. It was shown in Ref.~\cite{Griniasty88}
that for $\nu>1$ all one-electron states are localized. In this
regime, the frequency of the potential oscillations diverge in the
thermodynamic limit, resulting in a fairly uncorrelated sequence of
the on-site energies. Because of that, this case is usually named as
the {\it pseudo-random regime}. Oppositely, it was demonstrated that
the range $0 < \nu < 1$ supports a phase of extended states in the
neighborhood of the band center.~\cite{Sarma88Sarma90} In contrast
to the {\it pseudo-random regime}, it is the average wavelength of
the potential oscillations that diverges in the thermodynamic limit.
Because of the slower divergence of the oscillation wavelength, this
regime is usually named as {\it weakly aperiodic}. The uniform case
is recovered at $\nu=0$. Recently, the effect of this kind of
aperiodicity on the phonon and magnon $1D$ modes has  attracted a
renewed interest.~\cite{deMoura2006,chicomag2007}

The time-dependent Schr\"{o}dinger equation governs the dynamics of
the Wannier amplitudes,~\cite{Kramer93}
\begin{eqnarray}
i\,\dot{\psi}_{\bm m}&=& \left(\epsilon_{\bm m}+ \bm{U}\cdot
\bm{m}\right)\psi_{\bm m}\nonumber \\
&+&\left( \psi_{{\bm m}+{\bm e}_x}+ \psi_{{\bm m}-{\bm e}_x}+
\psi_{{\bm m}+{\bm e}_y}+\psi_{{\bm m}-{\bm e}_y} \right) \ ,
\label{Schrodinger}
\end{eqnarray}
where the Planck constant $\hbar$ is set to unity.

We solve numerically Eq.~(\ref{Schrodinger}) to study the time
evolution of an initially Gaussian wave-packet centered at site
${\bm m}_0$,
\begin{eqnarray}
\psi_{\bm m}(t=0)=A\exp\left[-\,\frac{(\bm{m}-\bm{m}_0)^2}{4\Delta^2}\right]\ ,
\label{Gaussian}
\end{eqnarray}
where $A$ is the normalization constant and we set $\Delta=1$ hereafter. Once
Eq.~(\ref{Schrodinger}) is solved for the initial condition~(\ref{Gaussian}), we
compute the projection of the mean position of the wave-packet (centroid) and
the average velocity  $v(t)$ along the field direction
\begin{eqnarray}
R(t)&=& \frac{1}{\sqrt{2}}\, \big[m_x(t)+m_y(t)\big], \nonumber \\
v(t)&=&-2\sum_{\bm m}\psi_{\bm m} \big[\psi_{{\bm m}+{\bm e}_x}^{*}
+\psi_{{\bm m}+{\bm e}_y}^{*}\big]\ ,
\label{tools1}
\end{eqnarray}
as well as the spread of the wave-function (square root of the mean-square
displacement)
\begin{equation}
\sigma(t)=\sqrt{\sum_{\bm m} \big[\bm{m}-\bm{m}(t)\big]^2 \,  |\psi_{\bm
m}(t)|^2} \label{tools2}
\end{equation}
where $\bm{m}(t)=\sum_{\bm m} \bm{m} \, |\psi_{\bm m}(t)|^2$.

In addition, we use the exact diagonalization of the
Hamiltonian~(\ref{hamiltonian}) in the absence of the electric field
($U=0$)  to obtain the eigenvalues and eigenvectors, and to
calculate the {\it dc} conductance $G(E)$ as a function of energy
$E$. For noninteracting electrons and within the linear response
theory at $T = 0\,$K, $G(E)$ is given by the Kubo-Greenwood
formula~\cite{tit} (in units of $e^2/\hbar$)
\begin{eqnarray}
G(E) &=& \frac{2\pi}{\Omega} \sum_{a,b}|\langle \,a\mid x{ H}-{
H}x\mid b\,\rangle|^2 \nonumber \\
&\times& \delta(E-E_a)\delta(E-E_b) \ .
\label{condu}
\end{eqnarray}
Here, the polarization is taken in the direction of the $x$-axis,
$\Omega$ is the area of the system, $x$ is the position operator,
and the indices $a$ and $b$ label the eigenstates. We also compute
the participation number of a normalized eigenstate as follows
\begin{equation}
P(E) =
\Big[\sum_{\bm m}|\psi_{\bm m}(E)|^4\Big]^{-1}\ .
\label{participation}
\end{equation}
For extended states, $P(E)$ is expected to scale linearly with the
number of sites. More specifically, we will focus on the {\it dc\/}
conductance and the participation number near the band center,
averaged over a narrow energy stripe at the band center $[-W_c/2,W_c/2]$:
\begin{subequations}
\begin{equation}
G=\frac{1}{N_{E}}\sum_{|E|<W_c/2}G(E) \ , \label{G_averaged}
\end{equation}
\begin{equation}
\xi =\frac{1}{N_{E}} \sum_{|E|<W_c/2}P(E)\ , \label{P_averaged}
\end{equation}
\end{subequations}
where $N_E$ is the number of eigenstates within the stripe.

\section{Results and Discussion}
\label{Results}

In all our numerical simulations, we set the parameters $V$ and
$\pi \alpha$ of the aperiodic potential to $V = 3$ and $\pi\alpha =
1$, while considering the exponents $\nu_x$ and $\nu_y$ as
variables. When calculating $G(E)$, Eq.~(\ref{condu}), the
$\delta$-functions were replaced by $\delta(E)=(1/\Delta
E)\,\Theta(\Delta E/2-|E|)$, with $\Delta E=0.1$, $\Theta$ being the
Heaviside step-function. The magnitude of the energy stripe in
Eqs.~(\ref{G_averaged}) and~(\ref{P_averaged}) was chosen $W_c=0.2$.

\subsection{Zero-field electronic states}

\begin{figure}[ht]
\begin{center}
\begin{tabular}{cc}
\includegraphics*[width=42mm,clip]{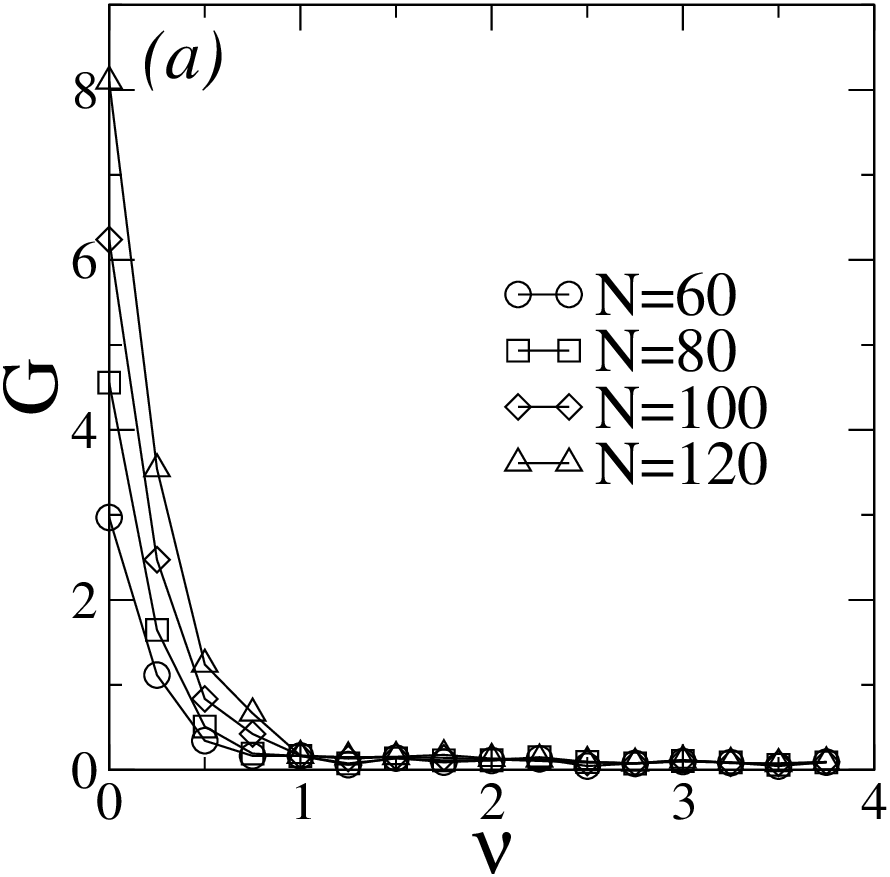} &
\includegraphics*[width=46mm,clip]{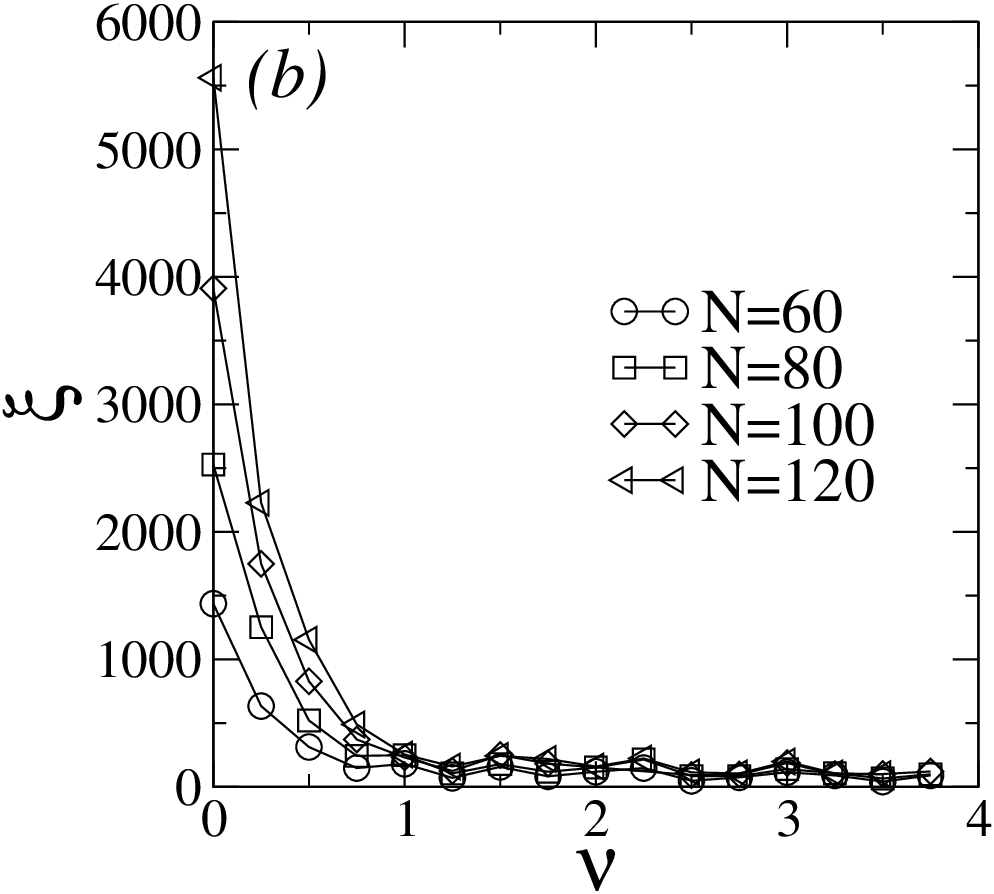}
\end{tabular}
\caption{(a)~Averaged {\it dc} conductance $G$,
Eq.~(\ref{G_averaged}), and (b)~averaged participation number
$\xi$, Eq.(\ref{P_averaged}), versus the exponent $\nu=\nu_x=\nu_y$
calculated for different system sizes $N$. The average was performed
considering a narrow range of potential amplitudes ($\Delta V=0.05$)
to reduce statistical fluctuations.}
\label{fig1}
\end{center}
\end{figure}

Firstly, we discuss the character of the electronic states of the
model under study. Figure~\ref{fig1}(a) displays the averaged
conductance $G$ at the band center ($E=0$) as a function of
$\nu\equiv \nu_x=\nu_y$, obtained after numerical diagonalization of
the Hamiltonian~(\ref{hamiltonian}) in the absence of the external
electric field ($U=0$). To avoid strong fluctuations of the
conductance related to particular realizations of the aperiodic
potential, we further averaged the results over a narrow range of
potential amplitudes ($\Delta V=0.05$). As we observe from
Fig.~\ref{fig1}(a), for a slowly varying aperiodic site potential
($\nu_x, \nu_y < 1$) the conductance grows on increasing the system
size. This is a signature of macroscopic conductance in the
thermodynamic limit. In contrast, the conductance is almost null for
a pseudo-random potential ($\nu_x, \nu_y > 1$), in agreement with
previous calculations for localized eigenmodes.~\cite{tit} A similar
trend is found in the behavior of the averaged participation number
$\xi$, as shown in Fig.~\ref{fig1}(b). For weakly aperiodic
potentials, $\xi$ grows with the system size, while it becomes size
independent for strongly aperiodic (pseudo-random) systems.

\begin{figure}[ht]
\begin{center}
\includegraphics*[width=55mm,clip]{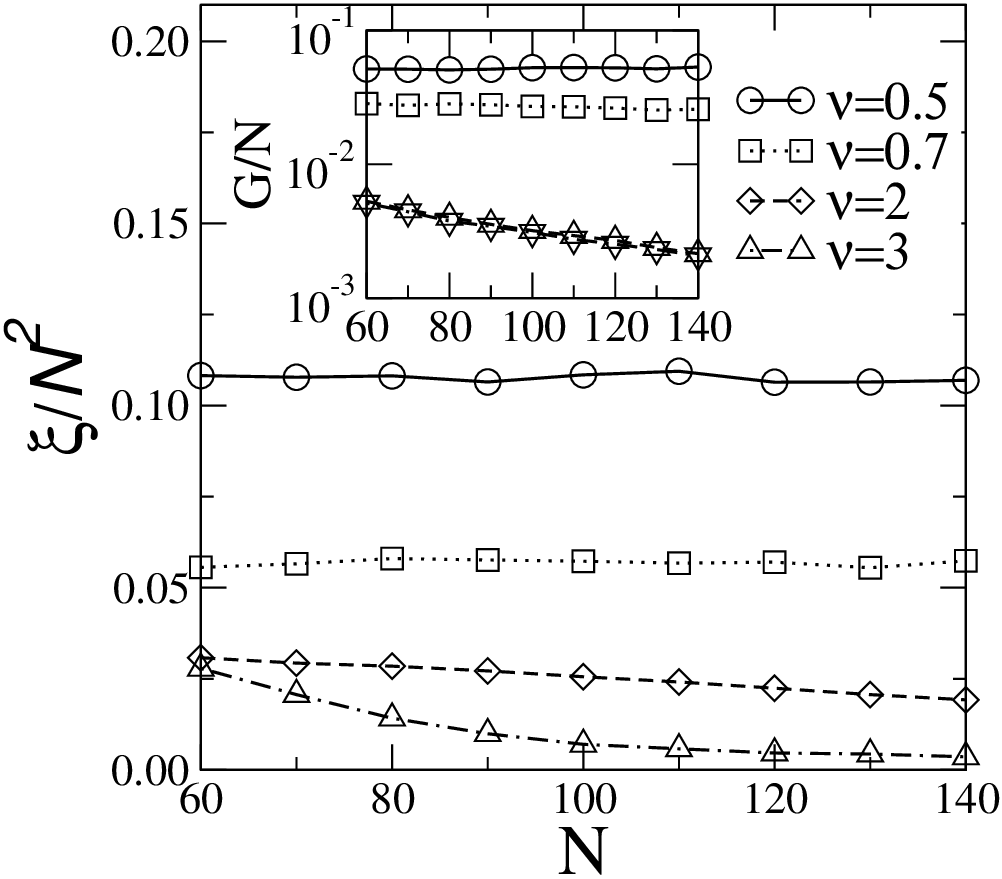}
\caption{Finite size scaling of the normalized participation number
$\xi/N^2$ for the states in the vicinity of the band center calculated for various values
of the exponent $\nu$ (shown in the legend). The inset presents the
finite-size scaling of the normalized conductance $G/N$. The different
character of the finite-size scaling of both quantities on increasing the
exponent $\nu$ signals the change in the nature of
electronic states in the band center from delocalized ($\nu < 1$) to localized ones
($\nu > 1$).
%the normalized participation $\xi/N^2$ and conductance $G/N$ are
%size independent in the regime of weak aperiodicity pointing to
%extended states and macroscopic conductance in the thermodynamic
%limit. For strong aperiodicities these quantities vanish for large
%$N$, in agreement with the usual scaling theory of localization.}
} \label{fig2}
\end{center}
\end{figure}

The correlations found in the behavior of the averaged conductance
$G$ and the participation number $\xi$ as a function of the exponent
$\nu$ indicate that the model under study undergoes a
metal-insulator transition associated with a change in the nature of
the one-electron eigenstates. To give further support to this
finding, we performed a finite-size scaling analysis of these two
quantities. In Fig.~\ref{fig2} we plotted the size dependence of the
normalized participation number $\xi/N^2$ for various values of the
exponent $\nu$. As it can be noticed,  $\xi/N^2$ does not depend of
$N$ in the regime of weak aperiodicity ($\nu < 1$), i.e., the
participation number itself $\xi$ is proportional to the total
number of sites $N^2$. This is a clear indication that the
eigenstates  at the band center are delocalized over a finite
fraction of the lattice. On the contrary, for strong aperiodic
potentials ($\nu
> 1$), $\xi/N^2$ decreases with the system size, signaling the
wave-function localization. The size dependence of the normalized
conductance $G/N$ in the center of the band, shown in the inset of
Fig.~\ref{fig2}, corroborates the conclusion above. In the slowly
varying aperiodic regime $G \propto N$ leading to a macroscopic
transport supported by extended states. Such a behavior is expected
for a regular square lattice, where the conductance at the band
center should be proportional to the number of conducting channels,
i.e., to the linear size of the lattice $N$. In the limit of
strongly aperiodic lattices, the normalized conductance decreases
with the system size, thus indicating the absence of macroscopic
transport in the thermodynamic limit. This is in agreement with the
standard scaling theory of Anderson localization.~\cite{abrahams}

\subsection{Dynamics}

We also study the time evolution of a wave-packet initially
localized in the center of the lattice, ${\bf m}_0 = (N/2,N/2)$, to
uncover the character of the electronic wave-packet dynamics in the
present model system. We start our analysis analyzing the
wave-packet dynamics in the absence of the external field ($U = 0$).
We numerically integrate the wave-equation~(\ref{Schrodinger}) and
calculate the wave-packet spread $\sigma(t)$ until it reaches a
stationary value as a result of multiple reflections of the electron
on the lattice boundaries. The extended states that appear in this
model emerge in the regime at which the potential has a diverging
wavelength in the thermodynamic limit. These are typically non
scattered modes thus leading to a ballistic wave-packet spread,
$\sigma(t) \propto t$. This means that, as long as $\sigma(t)
\leq N$, the data points calculated for different system sizes
$N$ should collapse into a unique straight line after re-scaling
$\sigma(t) \to \sigma(t)/N$ and $t \to t/N$. In the localized
regime, such a collapse is not expected. We further calculated the
time-dependent participation second moment  $\xi(t)=1/\sum_{\bm m}
|\psi_{\bm m}(t)|^4$. In general, $\xi(t)$ scales linearly with the
number of sites whenever the system supports extended states.

\begin{figure}[ht]
\centerline{\includegraphics[width=48mm,clip]{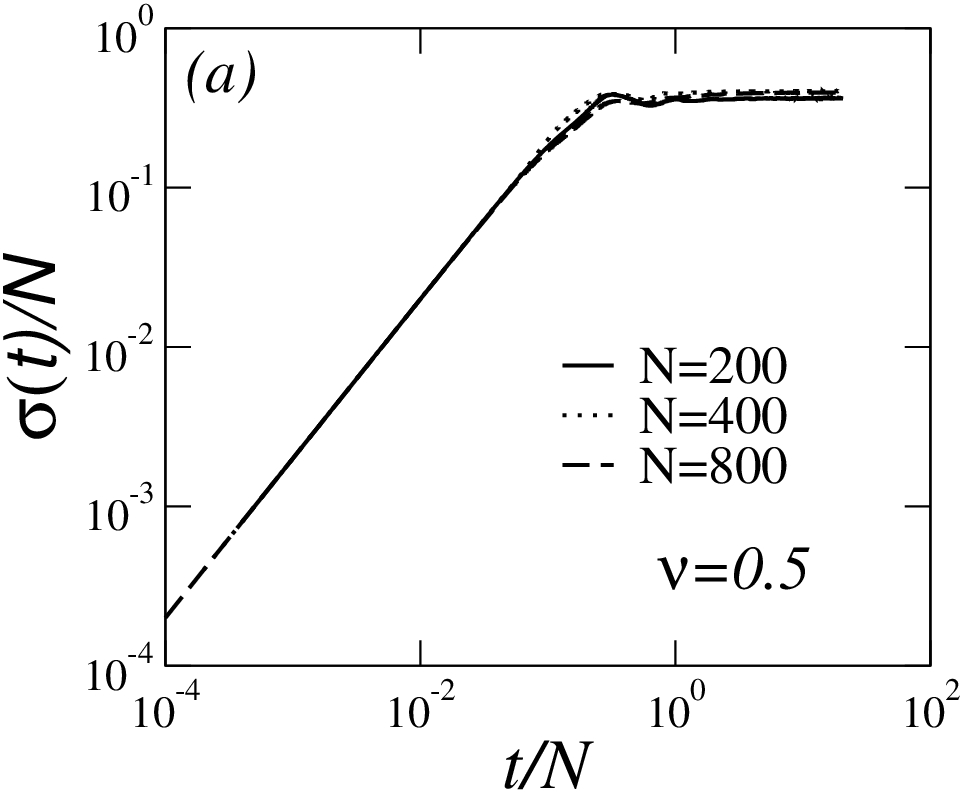}
\includegraphics[width=48mm,clip]{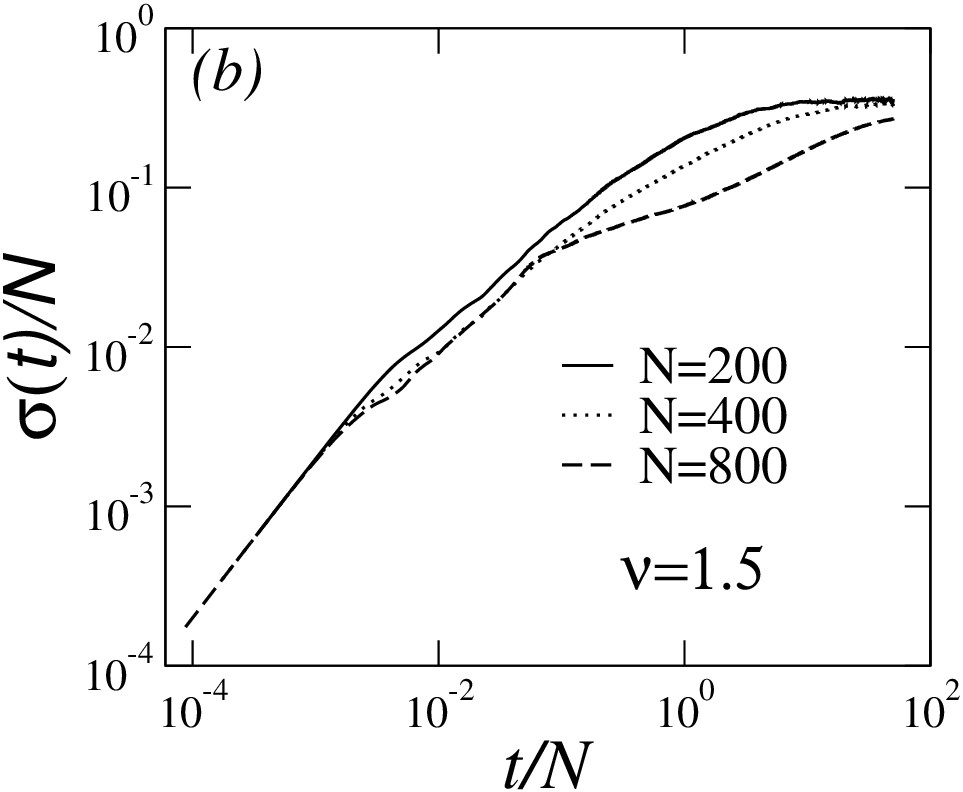}}
\centerline{\includegraphics[width=48mm,clip]{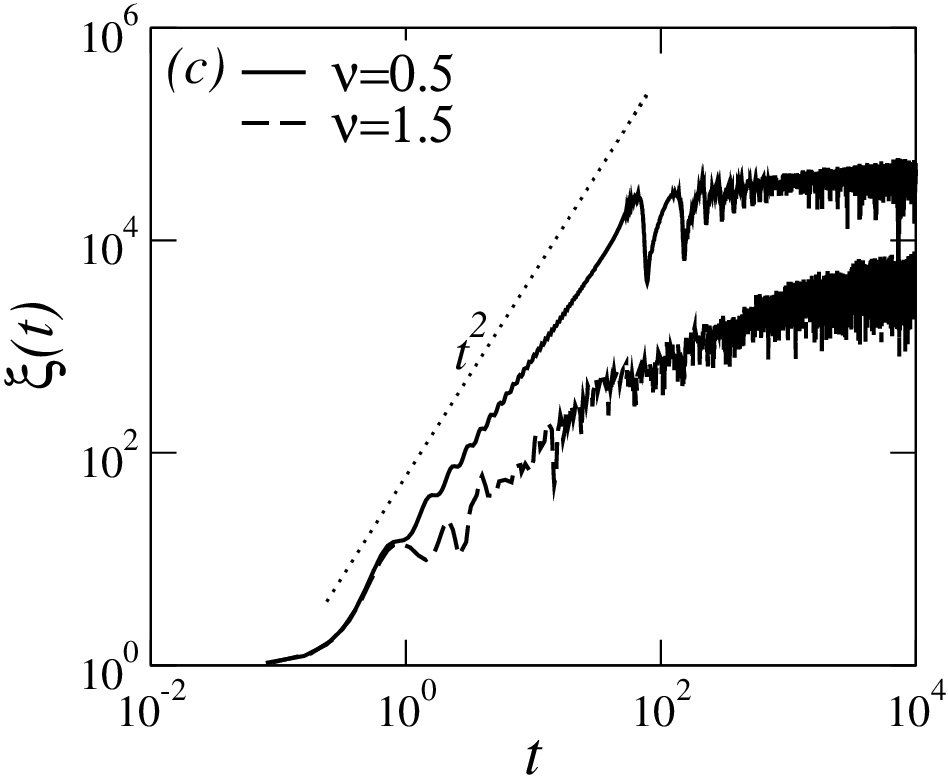}}
\caption{The normalized wave-packet spread $\sigma(t)/N$,
Eq.~(\ref{tools2}), as a function of the re-scaled time $t/N$
computed for lattices of $N \times N = 200\times 200$ to $800\times
800$ sites at zero bias ($U = 0$). (a) A slowly varying (weakly
aperiodic) site potential ($\nu_x = \nu_y = \nu = 0.5$). The
collapse of all curves into a unique curve until $\sigma(t) \leq
N$ indicates the presence of non scattered extended states. (b) A
pseudo-random potential ($\nu_x = \nu_y = \nu = 1.5$). A similar
collapse is absent, signalling localization. (c) Time dependent
participation $\xi(t)$ number versus time $t$ computed for $N \times
N = 1600\times 1600$ and $\nu_x = \nu_y = \nu = 0.5$ (solid line)
and $\nu_x = \nu_y = \nu = 1.5$ (dashed line). In agreement with the
wave-packet spread calculations, the participation number spreads
ballistically ($\xi(t)\propto t^2$) for $\nu<1$ indicating the
presence of non scattered modes.} \label{fig3}
\end{figure}

In Fig.~\ref{fig3}, we plot the results of our calculations. For a
weakly aperiodic site potential ($\nu_x = \nu_y = 0.5$), we found
the above mentioned collapse of all curves into a unique straight
line for $\sigma(t) \leq N$ [see Fig.~\ref{fig3}(a)]. This allow
us to conclude that the wave-packet motion in this case has a
ballistic nature, signaling the presence of a phase of delocalized
non scattered states. In the case of a pseudo-random site potential
($\nu_x = \nu_y = 1.5$), such a collapse is seen only for $\sigma(t)
\ll N$ [see Fig.~\ref{fig3}(b)]: the wave-packet motion is ballistic
whenever $\sigma(t)$ is smaller than the largest localization
length. For long times, a collapse is absent [see
Fig.~\ref{fig3}(b)], indicating the wave-packet localization. In
Fig.~\ref{fig3}(c), we show results for the time-dependent
participation second moment for $\nu_x = \nu_y = 0.5$  (solid line)
and $\nu_x = \nu_y = 1.5$ (dashed line). For $\nu<1$, the
participation dynamics show a ballistic behavior  until the
wave-packet  reaches the lattice boundaries. For $\nu>1$ a slower
dynamics was obtained. These results are in perfect agreement with
the wave-packet spread calculations,  indicating the presence of non
scattered modes in the initial wave-packet only in the regime of
weakly aperiodic potentials ($\nu<1$). These data corroborate our
statement about the Anderson transition for the underlined $2D$
aperiodic model, which we claimed in the previous section on the
basis of the finite size scaling of the conductance $G$ and
participation number $\xi$ for the states in the vicinity of the
band center.

\begin{figure}[ht]
\centerline{\includegraphics[width=80mm,clip]{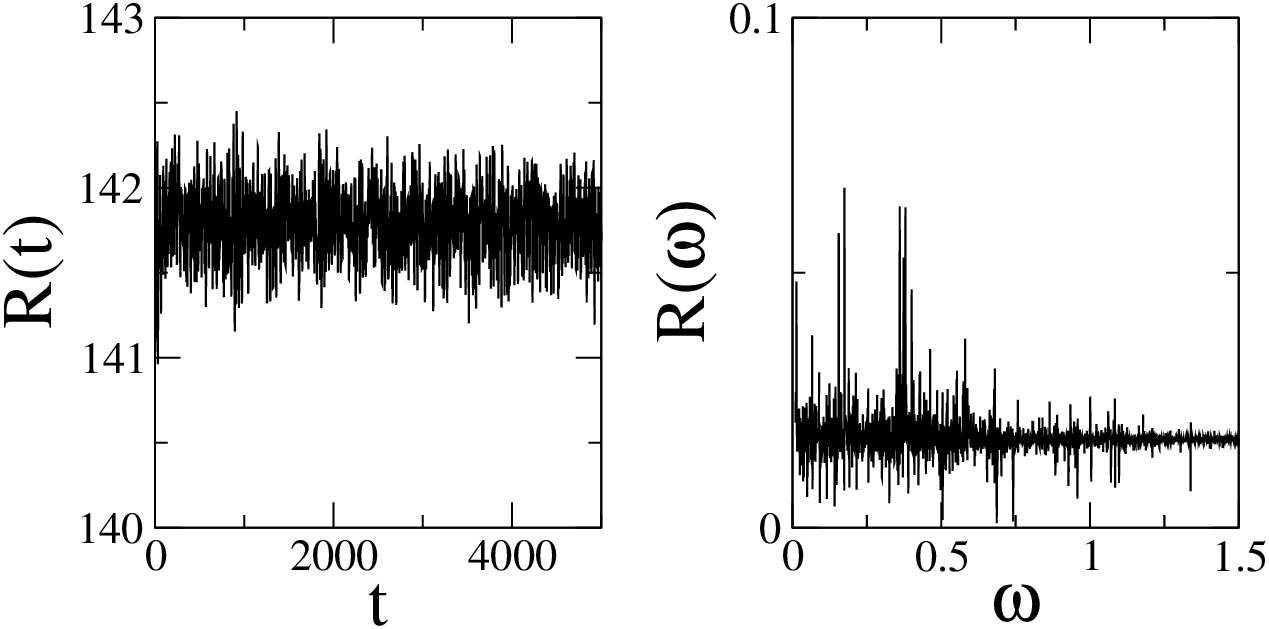}}
\caption{Bias driven dynamics of the centroid $R(t)$,
Eq.~(\ref{tools1}), in a pseudo-random potential ($\nu_x=\nu_y=
1.5$) with $N \times N=200 \times 200$ sites and bias magnitude
$U=0.5$. No signature of coherent Bloch oscillations is seen in
this case. The Fourier transform $\tilde{R}(\omega)$ of the centroid
$R(t)$ is broad, supporting  the absence of a typical oscillation
frequency in $R(t)$.} \label{fig4}
\end{figure}
\begin{figure}[ht]
\begin{center}
\begin{tabular}{cc}
\includegraphics*[width=41mm,clip]{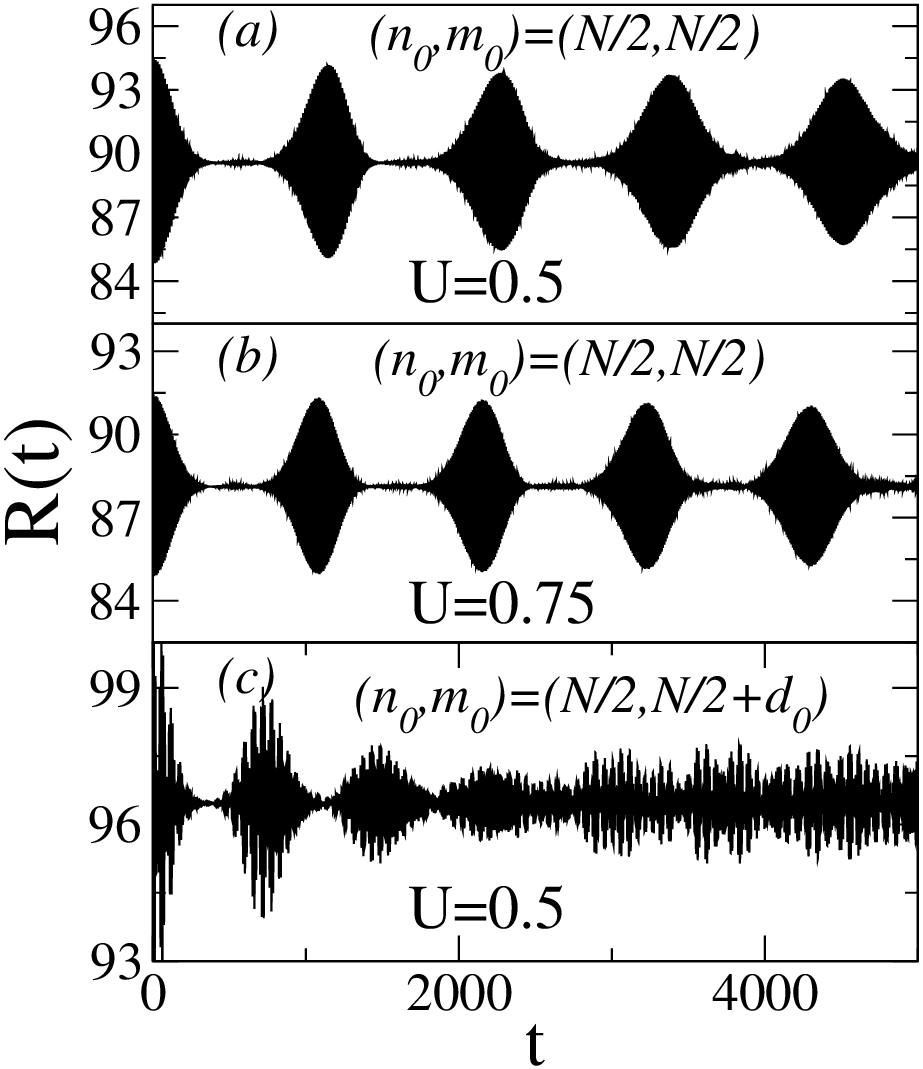} &
\includegraphics*[width=44mm,clip]{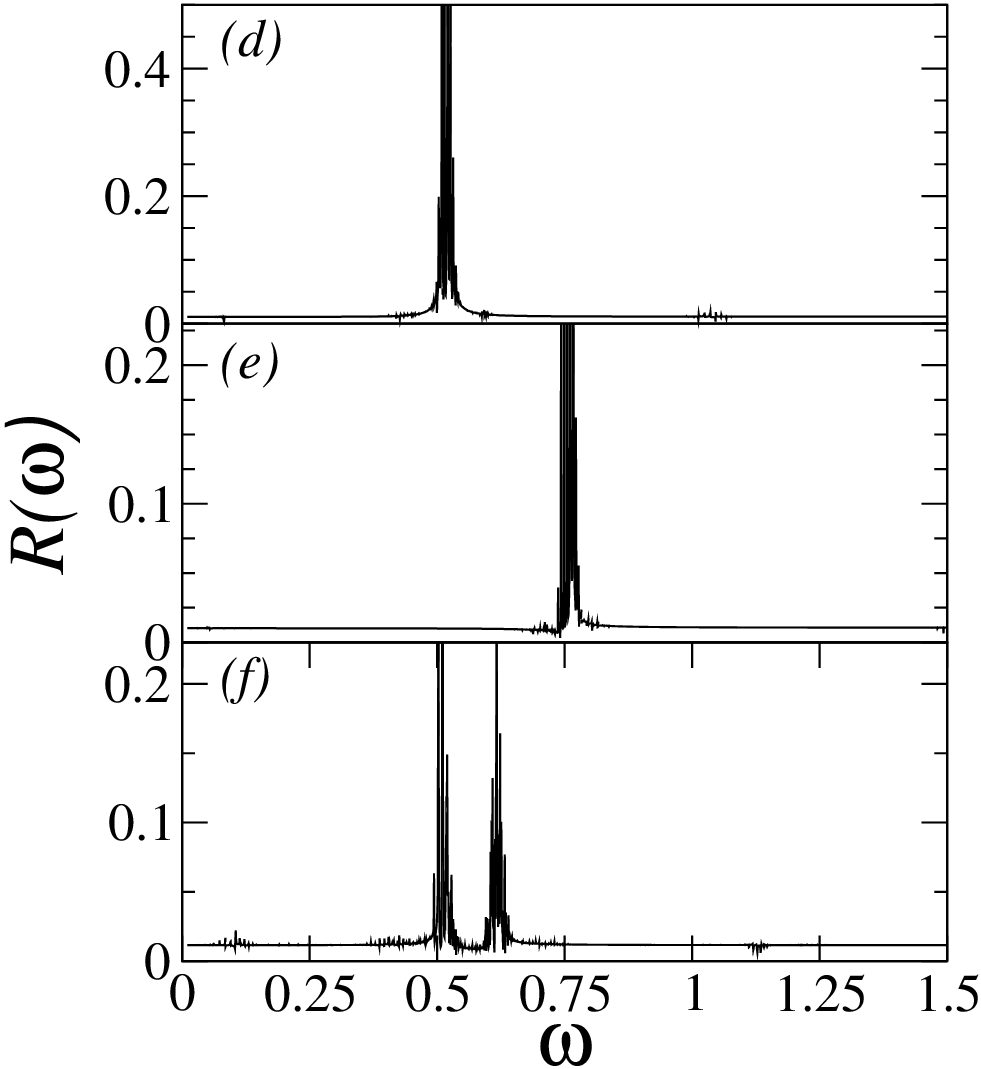}
\end{tabular}
\caption{Bias driven dynamics of the centroid $R(t)$ in a slowly
varying (weakly aperiodic) potential ($\nu_x = \nu_y = 0.5$) for two
magnitudes of the bias: (a)~$U=0.5$ and (b)~$U=0.75$. The initial
location of the wave-packet is $(n_0,m_0)=(N/2,N/2)$. (d)~and
(e)~Fourier transforms $\tilde{R}(\omega)$ of the centroids depicted
on panels (a) and (b), respectively. Note that $\tilde{R}(\omega)$
is peaked at a frequency $\omega \simeq U$. (c)~Same as in panel
(a), but for the initial condition $(n_0,m_0)=(N/2,N/2+d_0)$ with
$d_0=10$. (f)~Fourier transform of the centroid depicted on panel
(c). Note that here $\tilde{R}(\omega)$ exhibits a doublet
structure.} \label{fig5}
\end{center}
\end{figure}

\begin{figure}[ht]
\centerline{
\includegraphics*[width=70mm,clip]{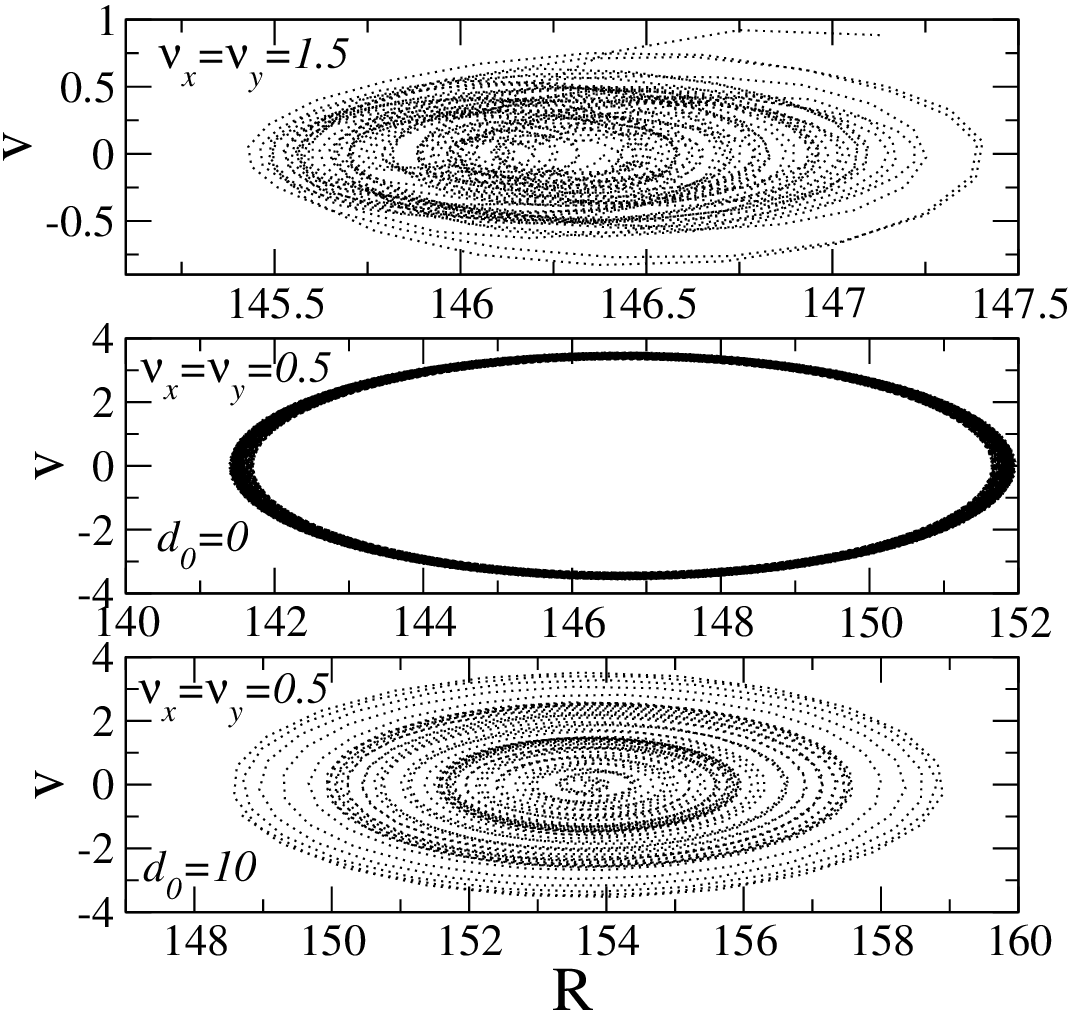}}
\caption{Phase diagrams velocity-vs-position computed for: (a)
$\nu_x = \nu_y = 1.5$, $U=0.5$ and the initial location of the
wave-packet $(n_0,m_0)=(N/2,N/2)$;  (b) $\nu_x = \nu_y = 0.5$,
$U=0.5$ and the initial location of the wave-packet
$(n_0,m_0)=(N/2,N/2)$; (c) the same as in (b) except the initial
location of the wave-packet $(n_0,m_0)=(N/2,N/2+d_0)$ with $d_0=10$.
For $\nu>1$ the velocity-vs-position diagram exhibits incoherent
orbits. For $\nu<1$, the Bloch-like oscillations promotes a coherent
dynamics through the phase space. In panel (c), the amplitude of the
oscillations varies faster than in panel (b) due to the doublet
structure generated by the anisotropic gradient of the local
potential.}
\label{fig6}
\end{figure}

Now, we turn to the wave-packet dynamics of an electron subjected to
a uniform electric field ($U \ne 0$). It is well known that in
disorder-free systems, a uniform electric field causes dynamic
localization of the electron and gives rise to an oscillatory motion
of the wave-packet, the so-called Bloch
oscillations.~\cite{Bloch28,Zener34} The size of the segment over
which the electron oscillates and the period of the oscillations are
estimated from semiclassical arguments to be $L_F = aW/U$ and
$\tau_B=2\pi/U$, respectively,~\cite{Ashcroft} where $W$ is the
width of the Bloch band.

Firstly, we compute the wave-packet centroid $R(t)$ in a pseudo-random
potential ($\nu_x = \nu_y = 1.5$) of size $N \times N = 200 \times
200$ with the bias magnitude $U=0.5$.  As deduced from
Fig.~\ref{fig4}, there is no signature of Bloch oscillations in this
case. Coherent oscillations, which are present immediately after the initial
wave-packet is released, are quickly destroyed (not shown). The asymptotic
behavior of the centroid resembles a stochastic motion around some
mean position. The Fourier spectrum $\tilde{R}(\omega)$ of the
centroid, plotted in the right panel of Fig.~(\ref{fig4}), confirms
this claim. The spectrum is rather broad, with no
characteristic frequencies, indicating that $R(t)$ is similar to a
white noise signal. Thus, for $\nu_x,\nu_y > 1$ the system shows a
behavior similar to the standard Anderson model, with no signature
of coherent Bloch oscillations.

\begin{figure}[ht]
\centerline{
\includegraphics*[width=85mm,clip]{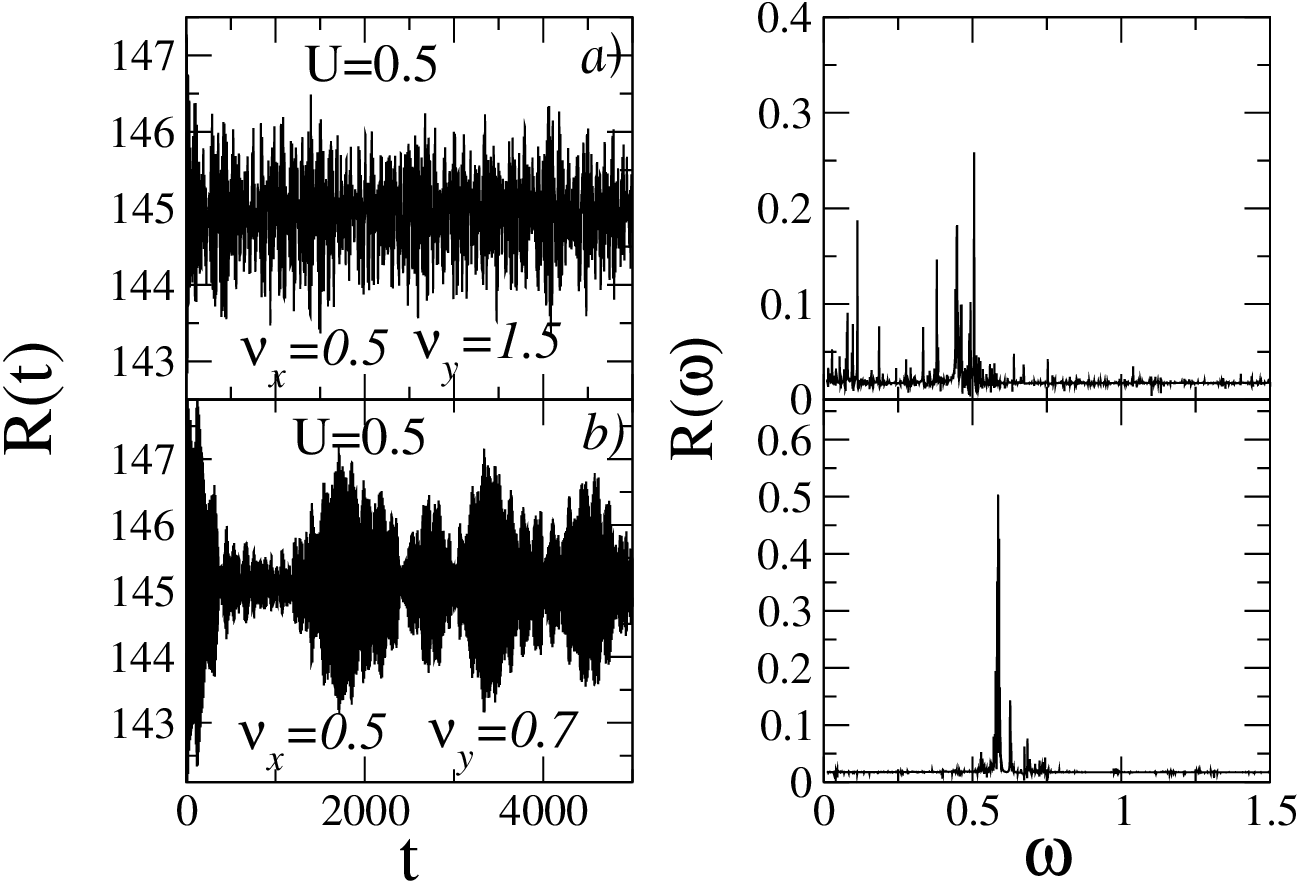}}
\caption{The dynamics of the centroid $R(t)$ and its Fourier transform
(left and right panel respectively) for $\nu_x=0.5$ and (a) $\nu_y=1.5$
and (b) $\nu_y=0.7$. For both $\nu_x$ and $\nu_y$ below $1$,
Bloch-oscillation  with a frequency close to the one predicted by
the semi-classical approach was obtained. However, in the case of
mixed pseudo-random and weakly aperiodic components, the pseudo-random
character for $\nu_y>1$ predominates leading to the Anderson localization
of the one-electron eigenstates and the consequent incoherent dynamics
of the wave-packet centroid. }
\label{fig7}
\end{figure}

In the case of a weakly aperiodic potential ($\nu_x=\nu_y=0.5$), the
centroid reveals an oscillatory amplitude-modulated pattern, as seen
in Fig.~\ref{fig5} (panels a, b, and c). Its Fourier transform
$\tilde{R}(\omega)$ displays a well-defined narrow peak, the
location of which is slightly deviated from the expected value
$\omega = U$ (panels d and e); this small shift is caused by the
local contribution to the external bias produced by the modulated
potential. Indeed, the bias at position $(n_0,m_0)$ is given by
${\bm U}_\text{eff} = (\delta\varepsilon_{n,m}/\delta n +
U,\delta\varepsilon_{n,m}/\delta m + U)$. The local contribution
shall be relevant whenever the potential gradient (in appropriate
units) is of the order of the external bias $U$. Therefore, the
frequency obtained in not exactly $\omega = U$ [Figs.~\ref{fig5}(a)
and~\ref{fig5}(b)]. The two peak structure depicted in
Fig.~\ref{fig5}(c) has its origin in the distinct contributions
given by the anisotropic gradient of the local potential to the
effective local bias. Once the initial wave-packet is located at a
non-symmetrical position of the lattice $(n_0,m_0)=(N/2,N/2+d_0)$
with $d_0=10$, the potential derivatives along the orthogonal
lattice directions give raise to distinct frequency shifts when the
electron is forced by a diagonal external field. We stress that such
frequency splitting is not found in $1D$ aperiodic  potentials,
being thus a specific feature of higher dimensional systems.

In Fig.~\ref{fig6}, we present the phase space plots, namely
velocity versus position .
All calculations were performed  using $N \times N = 200
\times 200$. Three representative cases are illustrated: (a)~$\nu_x
= \nu_y = 1.5$, $U=0.5$ with the initial location of the wave-packet
$(n_0,m_0)=(N/2,N/2)$;  (b)~$\nu_x = \nu_y = 0.5$, $U=0.5$ with the
initial location of the wave-packet  $(n_0,m_0)=(N/2,N/2)$; (c)~the
same as in (b) except by initial location of the wave-packet
$(n_0,m_0)=(N/2,N/2+d_0)$ with $d_0=10$. For all cases, the
integration time used was $t=1000$. We can see in (a) that the
velocity versus position diagram exhibits orbits  with no coherency,
as it was observed in the centroid dynamics (see Fig.~\ref{fig4}). In
the other cases, the diagrams show a coherent dynamics through the
phase space. In the case of bimodal Bloch oscillations [see
Fig.~\ref{fig6}(c)], the wave-packet orbits in phase space show a
pronounced breathing pattern. This breathing reflects the fact that
both  centroid and velocity display  amplitude-modulated envelopes.
The frequency of this amplitude-modulated envelope depends on the
difference between the two frequency peaks appearing the Bloch
oscillations spectrum [Fig.~\ref{fig5}(c)].

Before concluding, some words concerning the electronic dynamics in
systems with distinct degrees of aperiodicity along the orthogonal
lattice directions $\nu_x\not=\nu_y$. In Fig.\ref{fig7}, we show the
centroid $R(t)$ and its Fourier transform (left and right panel
respectively) for an initial Gaussian wave-packet located at the
center of the lattice with aperiodicity exponents $\nu_x=0.5$ and
(a) $\nu_y=1.5$ and (b) $\nu_y=0.7$. For both $\nu_x$ and $\nu_y$
below $1$, we can see oscillations with a frequency close to the
Bloch-oscillation frequency predicted by the semi-classical
approach. However, in the case of mixed pseudo-random and weakly
aperiodic components, the pseudo-random character for $\nu_y>1$
predominates leading to the Anderson localization of the
one-electron eigenstates and the consequent incoherent dynamics of
the wave-packet centroid.

\section{Summary and concluding remarks}
\label{Summary}

In this work, we studied numerically the dynamics of a single
electron wave-packet moving in a 2D square lattice with an
aperiodic  site potential $\epsilon_{\bm m}=V\cos{(\pi\alpha
m_x^{\nu_x})}\cos{(\pi\alpha m_y^{\nu_y})}$ in the presence of an
external uniform electric field. For fixed parameters $V$ and
$\pi\alpha$, the exponents $\nu_x$ and $\nu_y$ allow to control the
degree of aperiodicity.

We have discussed the character of the electronic states of the
model in the absence of an external electric field. Diagonalizing
numerically the Hamiltonian at zero electric field, we computed the
{\it dc} conductance and the participation number for various
systems sizes. After a finite-size scaling analysis of data close to
the band center, we found that, for a weakly aperiodic potential
($\nu_x,\nu_y < 1$), both quantities grow on increasing the system
size. The participation number is proportional to $N^2$ while the
conductance scales proportional to $N$. Such size dependencies are
consistent with the presence of extended states and macroscopic
transport in the thermodynamic limit.  In contrast, both
participation number and conductance reach a plateau upon increasing
the size for a pseudo-random potential ($\nu_x,\nu_y > 1$), which is
the typical behavior predicted by the usual scaling theory of
Anderson localization. Solving numerically the time-dependent
Schr\"{o}dinger equation, we found that an initially localized
wave-packet reveals a ballistic spreading associated with the
existence of a phase of extended non scattered states in the weakly
aperiodic limit ($\nu_x,\nu_y < 1$). In contrast, the spread of the
wave-packet is bounded for pseudo-random realization of the site
potential ($\nu_x,\nu_y > 1$). These results indicate that the
underlined $2D$ aperiodic model under study undergoes a
metal-insulator transition.

We have also investigated the interplay between the delocalization
effect, preserved by the weakly aperiodic structure, and the dynamic
localization, caused by an electric field acting on the system.  We
demonstrated that in weakly aperiodic potentials ($\nu_x,\nu_y <
1$), the applied uniform electric field promotes sustained Bloch
oscillations of the electronic wave-packet. The frequency of the
oscillations was shown to agree with the prediction of a
semi-classical approach.~\cite{Ashcroft} In contrast with the Bloch
oscillations in $1D$, the frequency spectrum of the centroid
time-evolution can exhibit a two-mode structure depending on the
initial position of the wave-packet. Such two-frequency dynamics
reflects the anisotropy of the potential landscape along the
principal lattice directions when the electron is forced to
oscillate by a diagonal field. In phase-space, this phenomenon is
signaled by quickly breathing coherent orbits.

\section{Acknowledgments}

Work at Alagoas was supported by CNPq-Rede Nanobioestruturas, CAPES
(Brazilian research agencies) and FAPEAL (Alagoas State agency).
Work at Madrid was supported by MEC (Project MOSAICO) and BSCH-UCM
(Project PR34/07-15916).

\end{document}